\documentstyle[preprint,pre,aps,epsfig]{revtex}
\newcommand{\BE}{\begin{equation}}
\newcommand{\EE}{\end{equation}}

\def\bq{\begin{equation}}
\def\eq{\end{equation}}

\begin{document}
\title{Front dynamics in reaction-diffusion systems with Levy flights: \\a fractional diffusion
approach
\footnote{Research sponsored by Oak Ridge National Laboratory,
managed by UT-Battelle, LLC, for
the U.S. Department of Energy under contract DE-AC05-00OR22725.}}
\author{D. del-Castillo-Negrete\thanks{e-mail: delcastillod@ornl.gov}\\
B. A. Carreras\\ V.~E. Lynch}
\address{Oak Ridge National Laboratory \\ Oak Ridge TN, 37831-8071}
\maketitle
\pacs{}
 
\begin{abstract}  
The use of reaction-diffusion models  rests on the key assumption that the underlying
diffusive process is Gaussian. However, a growing number of studies  have
pointed out the presence of   anomalous diffusion, and there is a need to understand the dynamics
of reactive systems in the presence of this type of non-Gaussian diffusion. 
Here  we present a study
of front dynamics in  reaction-diffusion systems where anomalous diffusion is due to 
asymmetric Levy flights. 
Our approach consists of  replacing the  Laplacian diffusion operator  by
a fractional diffusion operator,  whose fundamental solutions are Levy $\alpha$-stable
distributions.  
Numerical simulation of the fractional Fisher-Kolmogorov equation and 
analytical arguments show that anomalous diffusion leads to the exponential acceleration of
fronts and a universal power law decay, $x^{-\alpha}$, of the tail, where $\alpha$, the
index of the Levy distribution, is the order of the fractional derivative. 
\end{abstract} 
\pacs{PACS numbers: 82.40.Ck , 05.40.Fb, 02.50.Ey, 52.25.Fi}

%


%
%
%
%



Reaction-diffusion
models have found widespread applicability in many areas, including 
chemistry, biology, physics and engineering; see, for example, Refs.~\cite{reaction_diffusion} and
references therein.  
The simplest reaction-diffusion models 
are of the form 
\bq
\label{eq_1}
\partial_t \phi = \chi\,  \partial_x^2 \phi + F(\phi)\, ,
\eq
where $\chi$ is the diffusion constant, and  $F$ is a nonlinear function representing the reaction
kinetics.  Examples of particular interest include  the Fisher-Kolmogorov
equation for which $F=\gamma \phi(1-\phi)$ and the real Ginzburg-Landau equation for which
$F=\gamma \phi(1-\phi^2)$. 
The nontrivial dynamics of these systems arises from  the competition between
the reaction kinetics and diffusion.  

At a microscopic level, diffusion is the result of the random
motion of individual particles, and the  use of Laplacian  operators, $\partial_x^2 \phi$,  to model
it rests on the key assumption that this random motion is an stochastic Gaussian process.
However, a growing number of works have shown
the presence of anomalous diffusion processes for which the mean square variance,
$\langle \left[ x - \langle x \rangle \right]^2 \rangle \sim t^\gamma$,
grows faster ($\gamma>1$,  in the case of superdiffusion), or
slower ($\gamma<1$, in the case of subdiffusion) than in a Gaussian  
diffusion process \cite{anomalous_diffusion}. Accordingly, an  important open problem
 is to understand the dynamics of reaction-diffusion systems when
the assumption of Gaussian diffusion fails. 
This problem has a particular relevance to plasma physics, which was our original motivation to
carry out this study.  In particular, 
reaction-diffusion models with Gaussian diffusion  have been used to study the dynamics
of spatio-temporal propagating fronts in the transition to high confinement regimes in magnetically
confined plasmas \cite{LH_RD}. 
 However, these studies must incorporate
the fact that, in this type of plasmas, evidence
of non-Gaussian diffusion has been observed in  perturbative transport experiments
\cite{gentle}, in numerical simulation of three dimensional turbulence \cite{carreras}, and in
test-particle transport studies \cite{plasmas}.  

The origin of non-Gaussian diffusion can be traced back to the existence of
long-range correlations in the dynamics, or the presence of anomalously large particle
displacements described by broad probability distributions.  Here we are interested in the
second possibility. In particular, we focus on systems that exhibit anomalous diffusion caused by
Levy flights, for which the probability distribution of particle displacements, $p(\ell)$, is broad in
the sense that $\langle \ell^2\rangle=\infty$.  As it is well-known, for these kind of systems the
central limit theorem cannot be applied; and  as $N\rightarrow \infty$, the probability
distribution function  of $x =\sum^N_n \ell_n$, rather than being  Gaussian, is an
$\alpha$-stable Levy distribution
\cite{anomalous_diffusion,taqu}  

The majority of studies on anomalous diffusion due to Levy flights have focused on symmetric
processes  for which $p(\ell)=p(-\ell)$.  However, this is not always the case.  For example,
numerical studies of test particle transport in magnetized plasmas and in geophysical flows show
that these systems have a built-in asymmetry that gives rise to asymmetric transport 
\cite{plasmas,fluids}. 
Also, non-Gaussian asymmetric  processes are likely to have an important role in the
asymmetries observed in pulse propagation experiments in confined plasmas \cite{gentle}.
Motivated by this, we focus on anomalous diffusion processes that exhibit Levy flights in one
direction, say for $x>0$, but Gaussian behavior in the other direction, $x<0$.   The main goal of this
paper is to understand  how this asymmetry manifests in the
propagation of right-moving  ($x>0$) and left-moving  ($x<0$)  fronts in reaction-diffusion
systems. To study this problem we propose the following model:
\bq
\label{eq_2}
\partial_t \phi = \chi\,  _{a}D_x^\alpha \phi + F(\phi) \,  ,
\eq
\bq
\label{eq_3}
_{a}D_x^\alpha \, \phi =\frac{1}{\Gamma(2-\alpha)} \, \partial_x^2\, \int_{a}^x\,  
\frac{\phi(y)}{\left(x-y\right)^{\alpha-1}}  \,dy\, , 
\eq
where the Laplacian operator $\partial_x^2$ has
been replaced by 
$_{a}D_x^\alpha$, the  Riemann-Liouville, left fractional derivative of order $\alpha$, with
$1\leq \alpha <2$. Fractional calculus is a natural mathematical  generalization of standard
calculus \cite{fractional_text}, for example, 
$_{-\infty}D_x^{\alpha}\, e^{ikx} =(ik)^\alpha\,  e^{ikx}$,
and in recent years has been applied to a large class of problems in
the applied sciences; see 
Refs.~\cite{metzler,frac_diff_popular} and references therein. 

Previous studies of reaction-diffusion systems with anomalous diffusion include Ref.\cite{henry},
where reaction in the presence of subdiffusion was studied using   fractional
derivative operators in time.  In contrast, here we are interested in superdiffusion, which we model
with fractional operators in space. The interplay of bistable reaction processes and anomalous
diffusion caused by Levy flights was addressed in Ref.~\cite{zanette}. 
More recently,   superfast front
propagation in reactive systems with non-Gaussian  diffusion was discussed in
Ref.~\cite{vulpiani}.  The model studied in Ref.~\cite{vulpiani} consisted of a time-discrete
reaction system coupled to a superdiffusive Levy process described by an integral operator with an
algebraic decaying propagator.  In our model in Eq.~(\ref{eq_2}), we do not assume a
time-discrete reaction kinetics; and  through the use of fractional operators 
we use the exact Levy propagator.  In addition, we consider here the role of
asymmetric transport. 

The physical motivation for using the left-fractional derivative rests on the fact that 
for $F=0$, Eq.~(\ref{eq_2}) reduces to an asymmetric, space-fractional diffusion equation whose 
solution for a delta function initial condition,
$\phi(x,t=0)=\delta(x)$, in the infinity domain $x \in (-\infty, \infty)$   is 
\bq
\label{eq_5}
\phi(x,t)=\frac{1}{(\chi t)^{1/\alpha}}\, p_\alpha \left[ \frac{x}{(\chi t)^{1/\alpha}}\right] \, ,
\eq
where
\bq
\label{eq_6}
p_\alpha(\eta) = \frac{1}{2 \pi}\, \int_{-\infty}^{\infty} e^{i^\alpha k^\alpha + i k \eta}\, d k \, .
\eq
For $\alpha=2$,  Eq.~(\ref{eq_6}) is the Gaussian propagator, and Eq.~(\ref{eq_5}) is the
fundamental solution of the standard diffusion equation. 
However, for $1< \alpha<2$,  Eq.~(\ref{eq_6}) is an extremal, $\alpha$-stable Levy distribution, i.e.,
a distribution with maximum skewness. 
These distributions are the attractors of stochastic processes that exhibit Levy flights 
in only one direction,  they have algebraic asymptotic behavior at plus infinity,
$p_\alpha(\eta) \sim 1/\eta^{\alpha+1}$,
but decay exponentially at minus infinity \cite{taqu}.
In principle, one could add to Eq.~(\ref{eq_2}) a right-fractional
derivative  with a suitable weighting factor, and for $F=0$ obtain Levy
$\alpha$-stable distribution of arbitrary skewness \cite{space-frac}. 
Equation~(\ref{eq_5}) implies  that 
 $\langle x^n \rangle=(\chi t)^{n/\alpha}\, \int_{-r}^r
\eta^n p_\alpha(\eta) d \eta$ diverges as $r \rightarrow \infty$, for $n\ge \alpha$. However, in
physical applications (e.g., Ref.~\cite{fluids}), a finite-$r$ cut-off leads to the finite-size  scaling
$\langle x^n
\rangle
\sim t^{n/\alpha}$, which implies superdiffusive behavior with $\gamma=2/\alpha$.  

In the remainder of this paper, we present a numerical and analytical study of front solutions of
Eq.~(\ref{eq_2}) with a reaction dynamics 
of the Fisher-Kolmogorov type, $F=\gamma \phi(1-\phi)$, with
two fixed points, one stable, $\phi=1$, and the other unstable, $\phi=0$.
As it is usually done for initial value problems with fractional derivatives in time, and boundary
value problems in finite domains with fractional derivatives in space
\cite{metzler,fractional_text},  we regularize the singular behavior of the fractional operator  by
subtracting the value of $\phi(x)$ at the lower limit, namely
\bq
\label{eq_9}
\partial_t \phi = \chi\,  _{a}D_x^\alpha \left[\phi -\phi(a)\right] + \gamma \phi\, 
(1-\phi)\, ,
\eq
where $\phi(a)=\phi(x=a,t)$.
In an infinite domain, $a\rightarrow -\infty$, and Eq.~(\ref{eq_9}) reduces to
Eq.~(\ref{eq_2}) since the fractional derivative,  $_{-\infty}D_x^\alpha$, of a constant is zero. 
Here we consider the finite domain, $x\in(0,1)$, and take  $a=0$. 
The  numerical solutions were obtained by integrating Eq.~(\ref{eq_9})  with boundary conditions
$\phi(0)=1$ and $\phi'(0)=0$ for right-propagating fronts, and  $\phi'(0)=0$ and $\phi(1)=1$ for
left-propagating fronts.  We used a semi-implicit time advance with an up-wind finite difference
scheme.  The fractional operator was discretized using the 
Grunwald-Letnikov definition of the fractional derivative \cite{fractional_text}.  Details on  the
numerical method will be published elsewhere. 
 
\begin{figure}
\epsfig{figure=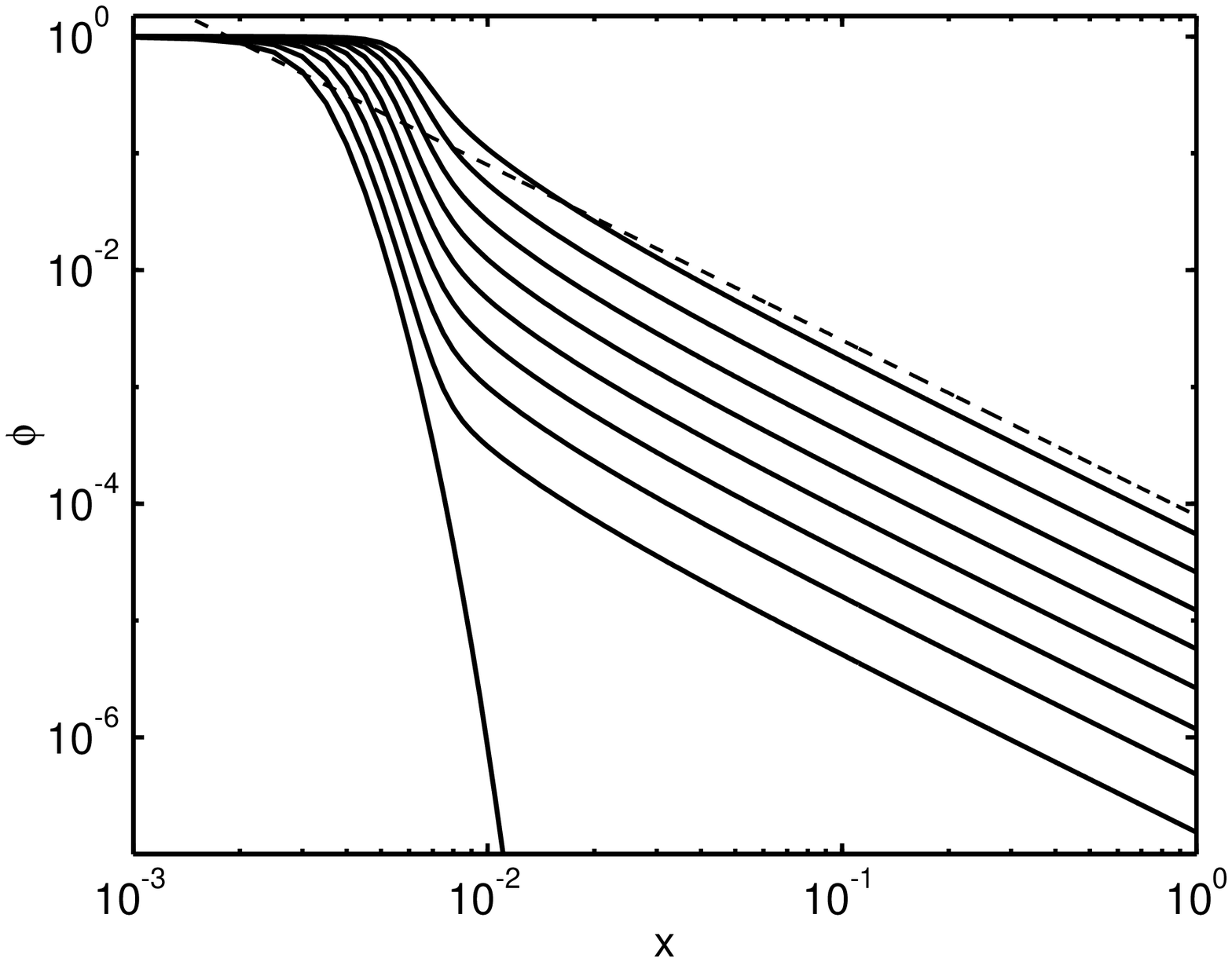,width=\columnwidth,angle=0}
\caption{Right propagating front profiles at successive times, obtained from a numerical integration
of the fractional Fisher-Kolmogorov Eq.~(\ref{eq_9}) with $\alpha=1.5$ and initial condition
$\phi(x,0)=\phi_0^r(x)$ in Eq.~(\ref{eq_10}). The dashed line has slope equal to $\alpha$. In
agreement  with the analytical result,  right-moving fronts develop an
algebraically decaying tail, $\phi\sim x^{-\alpha}$. }
\label{fig_1}
\end{figure}

\begin{figure}
\epsfig{figure=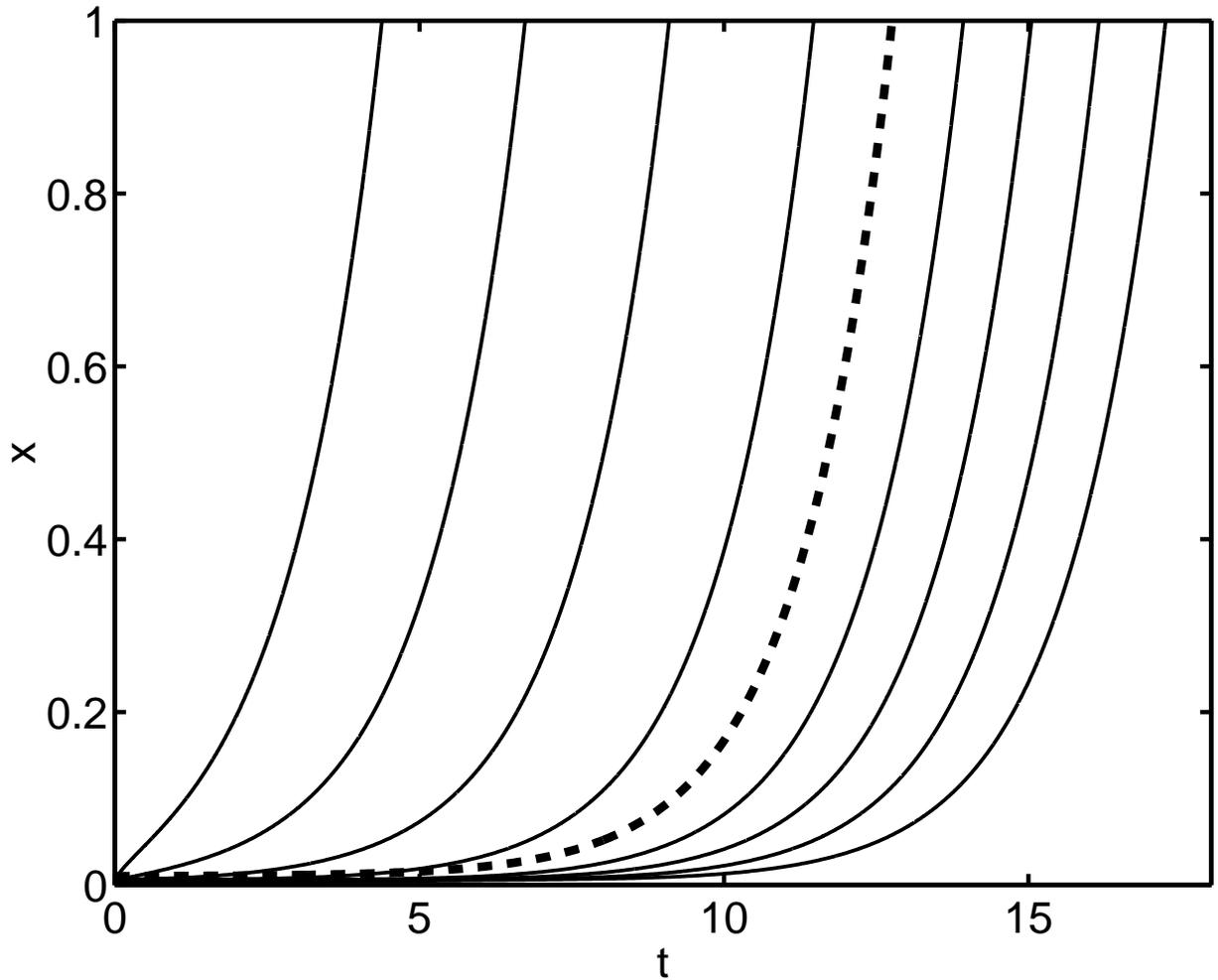,width=\columnwidth,angle=0}
\caption{Contour plot of the numerical solution $\phi(x,t)$  in Fig.~\ref{fig_1}. The curvature of
the iso-contours illustrates the exponential acceleration of fronts in the fractional
Fisher-Kolmogorov equation. The dashed line corresponds to the analytical scaling result 
$\phi \sim x^{-\alpha} e^{\gamma t}$.}
\label{fig_3}
\end{figure}

We consider two classes of initial conditions
\bq
\label{eq_10}
\phi^{r,l}_0(x)=\frac{1}{2}\left[ 1 \mp \tanh \left( \frac{x-x_0}{W} \right)\right] \, ,
\eq
where $\phi^{r}_0(x)$ takes the $-$ sign and corresponds to a right-propagating front, and 
$\phi^{l}_0(x)$ takes the $+$ sign and corresponds to a left-propagating front. 
Figure~\ref{fig_1} shows the time evolution of the front profile for 
$\phi^r_0(x)$ with $W=0.001$ and $x_0=0.003$, obtained from the direct numerical integration of
Eq.~(\ref{eq_9}) with $\gamma=1$ and $\chi=5\times 10^{-7}$. In this case, the front propagates
to the right and develops an algebraic decaying tail $\phi \sim x^{-\alpha}$.
The time evolution of the tail, $\phi(x=1,t)$, exhibits, for large $t$, exponential growth
$\phi \sim e^{\gamma t}$.  The acceleration of the front is evident in the space-time diagram in 
Fig.~\ref{fig_3} that shows a contour plot of $\phi(x,t)$.
Front acceleration  and 
algebraic decay of the tail was also observed in the model studied in Ref.~\cite{vulpiani}, but with
a different  exponent, namely $\phi \sim x^{-(\alpha+1)}$.

\begin{figure}
\epsfig{figure=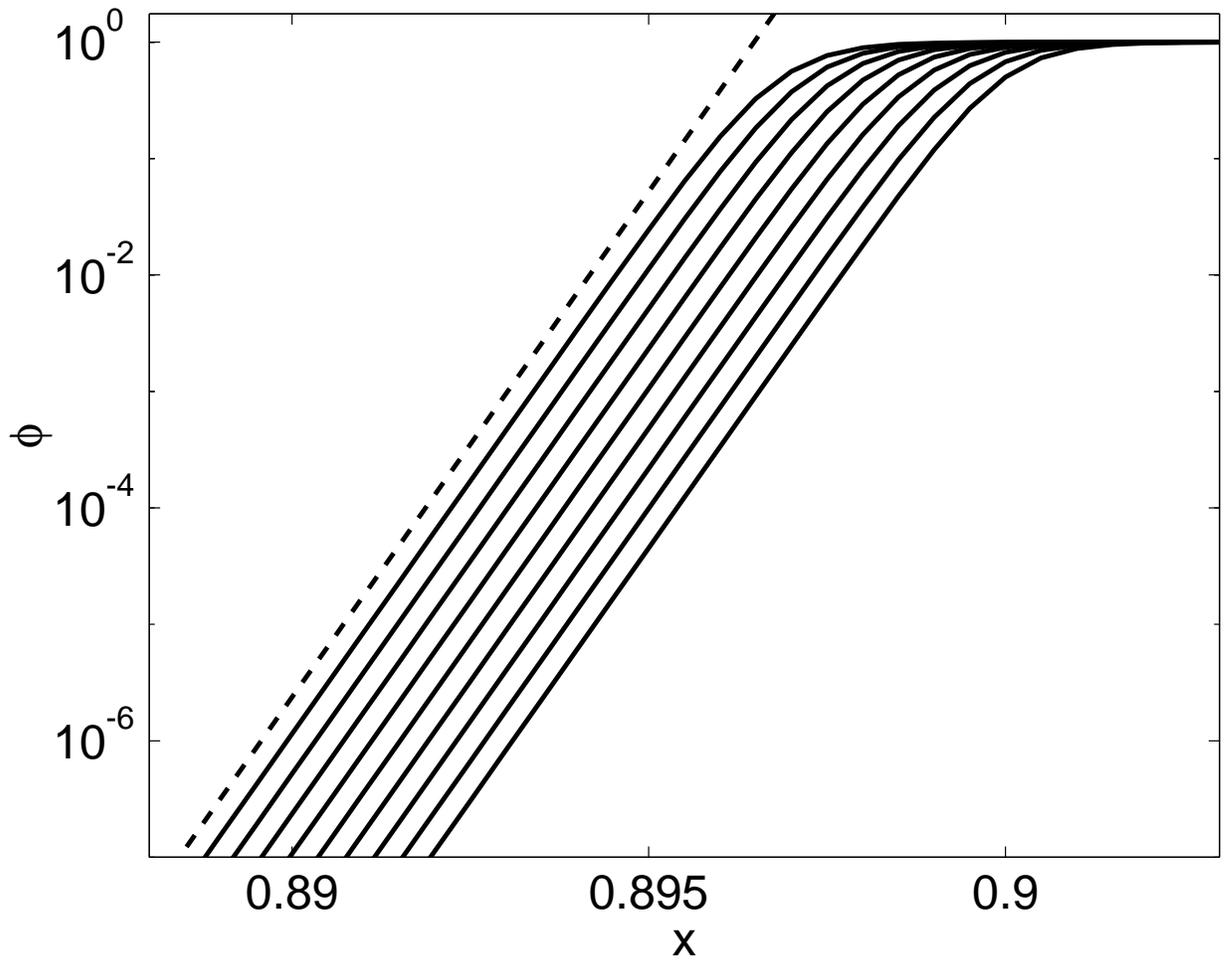,width=\columnwidth,angle=0}
\caption{Left-propagating front profiles at successive times obtained from a numerical integration
of the fractional Fisher-Kolmogorov Eq.~(\ref{eq_9}) with $\alpha=1.5$ and initial condition
$\phi(x,0)=\phi^l(x)$ in Eq.~(\ref{eq_10}). The dashed line has slope equal to $\lambda=2/W=2000$.
In agreement  with the analytical result in Eq.~(\ref{eq_18}), the front exhibits an exponential
decay  and has a speed of $c=5.22\times10^{-4}$.}
\label{fig_4}
\end{figure}

Left-propagating fronts  exhibit a more standard dynamics. 
In particular, Fig.~\ref{fig_4} shows the time evolution of the front profile for an initial
condition
$\phi^l_0(x)$ with  $W=0.001$, $\gamma=1$, and $\chi=5\times 10^{-7}$ (the same parameter
values used in Fig.~\ref{fig_1}) and $x_0=0.9$. 
In this case, the front exhibits an exponential decay and a
self-similar propagation with constant speed $c$. 
%

The numerical results presented above can be explained analytically using the leading-edge
approximation, extensively used in the study of fronts with Gaussian diffusion \cite{saarloos}. 
In this approximation,  in the limit $\phi\ll1$, the reaction kinetics
$F(\phi)$ is linearized around the unstable phase  leading to the linear fractional  equation:
\bq
\label{eq_11}
\partial_t \phi = \chi\,  _{-\infty}D_x^\alpha \phi \, +\,  \gamma \phi\, .
\eq
Substituting  $\phi=e^{\gamma t}\, \psi(x,t)$ into Eq.~(\ref{eq_11}) yields a fractional diffusion
equation for $\psi$ whose general solution is 
\bq
\label{eq_12}
\psi(x,t) = \int_{-\infty}^{\infty} p_\alpha (\eta)\,
\psi_0[x-(\chi t)^{1/\alpha}\eta\,]\, d\eta\, ,
\eq
where, as discussed before, 
$p_\alpha(\eta)$ given in Eq.~(\ref{eq_6}) is the extremal Levy $\alpha$-stable distribution that
is the Green's function of the asymmetric fractional diffusion equation. 
 For a front initial condition of the form
$\phi_0(x<0)= 1$  and $\phi_0(x>0)=e^{-\lambda x}$,
Eq.~(\ref{eq_12}) gives 
\bq
\label{eq_13}
\phi(x,t) = e^{\gamma t}\,  \int_{x (\chi t)^{-1/\alpha}}^{\infty}
p_\alpha(\eta)\, d \eta+
e^{-\lambda x+\gamma t}\, \int_{-\infty}^{x (\chi t)^{-1/\alpha}} 
e^{\lambda (\chi t)^{1/\alpha}\, \eta}
\,p_\alpha(\eta)\, d \eta	\, .
\eq

Before using this  solution to obtain the asymptotic behavior of fractional diffusion,
right-propagating fronts, it is instructive to consider the standard diffusion limit, $\alpha=2$. In
this case, the   Green's function is the Gaussian propagator $p_2(\eta)$,
and Eq.~(\ref{eq_13}) becomes
\bq
\label{eq_14}
\phi(x,t)= e^{-\lambda\, (x-ct)}\, P\left(\frac{x-2 \lambda \chi t}{\sqrt{2 \chi t}}\right) 
+ e^{\gamma t}\,\left[1-P \left(\frac{x}{\sqrt{2 \chi t}}\right) \right] \, ,
\eq
where $P(z)$ 
is the normal probability distribution function. 
Using the fact that $P(z\rightarrow \infty)=1$, we get from Eq.~(\ref{eq_14}) the 
asymptotic  behavior  $\phi(x,t) \sim e^{-\lambda (x-ct)}$.
Where the speed of the front, $c$, is related to the steepness of the front, 
$\lambda$, according to $c=\gamma/\lambda + \chi \lambda$, with the minimum front speed
$c_m=2 \sqrt{\gamma \chi}$ selected for sufficiently steep profiles
\cite{reaction_diffusion,saarloos}.

In the fractional case we consider a large, fixed $t$, and  $x (\chi t)^{-1/\alpha} \rightarrow
\infty$.  Introducing a cut-off $\Omega$ such that $1\ll\Omega< x (\chi t)^{-1/\alpha}$ in the
second integral in Eq.~(\ref{eq_13}),   integrating by parts,
and using the asymptotic expression of the Levy distribution, 
$p_\alpha(\eta) \sim 1/\eta^{\alpha+1}$ we get
\bq
\label{asymp}
\phi\sim \chi t e^{\gamma t}\, \left[\frac{x^{-\alpha}}{\alpha}+
 \frac{x^{-\alpha-1}}{\lambda}+ 
\frac{(1+\alpha)}{\lambda \left(\chi t\right)^{1+1/\alpha}}\, 
\int_{\Omega}^{x (\chi t)^{-1/\alpha}} 
\frac{e^{\lambda \left[(\chi t)^{1/\alpha} \eta-x\right]}}{\eta^{2+\alpha}}\, d \eta+\ldots\right]\, ,
\eq
where the dots denote terms of order $e^{-\lambda x}$. 
Since the integrand on the right hand side of Eq.~(\ref{asymp}) is bounded by $1/\eta^{2+\alpha}$,
the third term in this equation is at most of order $x^{-\alpha-1}$. Therefore, 
in agreement with the numerical result in Fig.~\ref{fig_1}, to leading order the
tail of right propagating fronts decays as $\phi\sim 1/x^\alpha$. 
The time-asymptotic dynamics for fixed, large $x$ can also be obtained from
Eq.~(\ref{eq_13}). In this case, a stationary phase approximation gives, in agreement with the
numerical results, $\phi \sim e^{\gamma t}$ to leading order, with
a correction of order $(\chi t)^{-1/\alpha}\, e^{\gamma t}$.

The  previously discussed asymptotic results can be summarized as
$\phi \sim x^{-\alpha}\, e^{\gamma t}$,
which, as Fig.~\ref{fig_3} shows, reproduces the numerical results. 
Solving the equation $\phi(x,t)=\phi_L$ for
$x$, for a fixed value  $\phi=\phi_L \in (0,1)$ gives the Lagrangian trajectory $x_L=x(t; \phi_L)$ of
the coordinate of a point in the front with concentration $\phi=\phi_L$. In the asymptotic limit
we have $x_L \sim e^{\gamma t/\alpha}$. That is, the Lagrangian velocity
$v_L=dx_L/dt$ of right-moving fronts  grows exponentially with time,
$v_L \sim e^{\gamma t /\alpha}$.

To conclude, consider the dynamics of left-moving fronts, like the one shown in  Fig.~\ref{fig_4}. 
In this case, the leading-edge description  is straightforward since these
fronts exhibit exponential tails and propagate at constant speed. Substituting 
 $\phi\sim \exp [\lambda ( x + ct) ]$ into Eq.~(\ref{eq_2}), we obtain the dispersion
relation $c=c(\lambda)$, with minimal front speed $c_{m}=c(\lambda_m)$ where
\bq
\label{eq_18}
c=\frac{\gamma}{\lambda}+\frac{\chi}{\lambda^{1-\alpha}}\, , \qquad
c_{m}=\alpha \chi^{1/\alpha}\, \left(\frac{\gamma}{\alpha-1}\right)^{(\alpha-1)/\alpha} \, .
\eq
Numerical results support the idea that for steep ($\lambda\geq \lambda_m$)  initial conditions
the front selects the minimum velocity $c_m$, whereas for wide 
($\lambda < \lambda_m$) initial conditions  the front speed depends on  the initial
condition according the first equation in (\ref{eq_18}).  As expected, in the limit $\alpha=2$
the well-known results of front propagation in the presence of Gaussian diffusion 
\cite{reaction_diffusion,saarloos} are recovered. 

 Summarizing, in this paper we have proposed the
use of fractional-diffusion operators to study front dynamics in reaction-diffusion systems with
non-Gaussian diffusion caused by asymmetric Levy flights.  Numerical and analytical results show
that right-moving fronts accelerate exponentially, and develop an  algebraic decaying  tail, 
$\phi \sim x^{-\alpha}\, e^{\gamma t}$.   Left-moving fronts have exponential decaying tails and
move at a constant speed given by  Eq.~(\ref{eq_18}). The results are general in the sense that they
are independent of the details of the reaction kinetics, provided it is of the ``pull" type with a
stable and an unstable phase.  Two areas of potential
application of the ideas presented here are Plasma Physics and Biology. 
In particular,  transport studies in three-dimensional pressure-driven plasma turbulence
\cite{carreras} and in drift waves \cite{plasmas} have shown evidence of anomalous diffusion and
non-Gaussian probability distributions of particle displacements which can be modeled using
fractional diffusion equations.  Also, fractional diffusion equations seem to 
capture important aspects of perturbative transport experiments  in fusion plasmas including 
nonlocal diffusion effects in the propagation of cold pulses (e.g., \cite{gentle}).  
On a parallel development, reaction-diffusion models with Gaussian diffusion
have  been used to study the turbulence-shear flow interaction in the L-H
transition in fusion plasmas (e.g., \cite{LH_RD}), and there is a pressing need to understand the 
role of non-Gaussian, non-local diffusion in the dynamics of L-H transition fronts. The results
presented here represent a first step in the study of this open problem. 
On the other hand, the fractional Fisher-Kolmogorov
equation discussed here shares some similarities with reaction equations with
integro-differential operators used in Biology to model long-range diffusion and spatial patterns
in neural firing \cite{reaction_diffusion}. In this regard, the algebraic decaying kernel in the
fractional diffusion operator might be useful for describing strongly non-local processes. 

We thank Angelo Vulpiani and Michael Menzinger for useful conversations.  This work was sponsored
by the Oak Ridge National Laboratory, managed by UT-Battelle, LLC, for the U.S. Department of
Energy under contract DE-AC05-00OR22725.

\nopagebreak
%

%




%
%

\end{document}